\documentclass[%
 reprint,
 superscriptaddress,
 nofootinbib, 
 nobibnotes,
 amsmath,amssymb,
 aps,
pdftex,
]{revtex4-1}

\usepackage{graphicx}
\usepackage{dcolumn}
\usepackage[hypertexnames=false]{hyperref}
\usepackage{enumerate}
\usepackage{bm}
\usepackage{mathrsfs}
\usepackage{comment}
\usepackage{url}
\usepackage{siunitx}

\usepackage[utf8x]{inputenc}
\usepackage[T1]{fontenc}

\newcommand{\unit}[1]{\,\rm{#1}}
\newcommand{\eq}[2][]{
	\begin{align#1}
		#2
	\end{align#1}
}

\newcommand{\myfig}[4][width=\linewidth]{
    \begin{figure}[tb]
        \centering
            \includegraphics[#1]{#2}    
        \caption{#3}                    
        \label{#4} 
    \end{figure}
}

\newcommand{\diff}{\mathrm{d}}
\newcommand{\iu}{\mathrm{i}}
\newcommand{\e}{\mathrm{e}}

\newcommand{\cross}{\times}
\newcommand{\figref}[1]{Fig.~\ref{#1}}
\newcommand{\tabref}[1]{Table~\ref{#1}}
\newcommand{\secref}[1]{Sec.~\ref{#1}}
\newcommand{\appref}[1]{Appendix~\ref{#1}}

\newcommand{\av}[1]{\left\langle #1 \right\rangle}
\newcommand{\hatilde}[1]{\hat{\tilde{#1}}}

\allowdisplaybreaks[4] 

\begin{document}


\title{High Speed Source Localization in Searches for Gravitational Waves from Compact Object Collisions}

\author{Takuya Tsutsui}
\affiliation{%
Research Center for the Early Universe (RESCEU), Graduate School of Science, The University of Tokyo, Tokyo 113-0033, Japan
}
\affiliation{%
Department of Physics, Graduate School of Science, The University of Tokyo, Tokyo 113-0033, Japan
}
\author{Kipp Cannon}%
\affiliation{%
Research Center for the Early Universe (RESCEU), Graduate School of Science, The University of Tokyo, Tokyo 113-0033, Japan
}
\author{Leo Tsukada}%
\affiliation{%
Research Center for the Early Universe (RESCEU), Graduate School of Science, The University of Tokyo, Tokyo 113-0033, Japan
}
\affiliation{%
Department of Physics, Graduate School of Science, The University of Tokyo, Tokyo 113-0033, Japan
}

\date{\today}

\begin{abstract}
Multi-messenger astronomy is of great interest.
The localization speed of gravitational wave sources is important for the success of electromagnetic follow-up.
Although current gravitational wave source localization methods take up to a few seconds, even that is not sufficient for some electromagnetic bands.
Therefore, one needs a more rapid localization method even if it is less accurate.
Building upon an Excess power method, we describe a new localization method for compact object collisions that produces posterior probability maps in only a few hundred milliseconds.
Some accuracy is lost, with the searched sky areas being approximately $10$ times larger.
We imagine this new technique playing a role in a hierarchical scheme where fast early location estimates are iteratively improved upon as better analyses complete on longer time scales.

\end{abstract}

\maketitle


\section{Introduction}

In August 17 2017, the Advanced LIGO~\cite{LIGO1, LIGO2} and the Advanced Virgo~\cite{Virgo} observed a gravitational wave (GW) from binary neutron star (BNS) merger, dubbed as GW170817~\cite{GW170817_observation}.
Then, many electromagnetic (EM) telescopes followed it to find the EM counterpart with multi-wavelength from radio wave to gamma ray~\cite{GW170817_multimessenger}.
By these observations, BNS merger was corroborated to be the origin of short gamma ray burst (sGRB), which had been discussed for a long time~\cite{CBC-GRB}.
The coordinated observation by different means of astronomical signals, for example, GW and EM wave is so-called multi-messenger astronomy.
GW170817 is one of the successful cases of multi-messenger astronomy.
By multi-wavelength observations, information of systems is much more increased.
Furthermore, the third observing run (O3) with improved sensitivity was done in 2019 and many observations with higher sensitivity have been already planned.
GW observation is expected to play a more important role in physics.

So far, all detected GW events are compact binary coalescence (CBC).
The GW signal stays in the sensitive band of ground-based detectors for minutes during the early inspiral phase, which the waveform is well known.
Thus if CBC signals are sufficiently loud, one can detect them before the merger by accumulating enough signal to noise ratio (SNR) to detect, which is called early warning~\cite{early_warning}.
This could bring scientific benefits for multi-messenger astronomy because one can prepare for transient events and observe precursor events.
For example, there are prompt optical flash from BNS~\cite{prompt_flash}, characteristic EM emission from tidal disruption of neutron star-black hole (NSBH) before merger~\cite{NSBH_EMemission}, resonant shattering of neutron star (NS) crusts~\cite{resonant_shattering}, NS magnetospheric interactions~\cite{magnetic_interaction_in_BNS}, and fast radio burst driven by black hole (BH) battery~\cite{FRB_BH_battery}.
There should be many such undiscovered events over multi-wavelength.

In an early warning context, location estimates can be iteratively improved.
There are currently two stages of refinement, BAYESTAR~\cite{bayestar, bayestar3d} which takes about $\SI{3}{s}$, and LALInference~\cite{LALInference} which takes hours to days.
However, there is a need for a still faster location estimate even at the expense of localization accuracy.
A rough location estimate available in $\mathcal{O}(\SI{100}{ms})$ could trigger the slewing of fast facilities like Cherenkov telescopes, allow better data retention decisions by low-frequency radio facilities, and it might even be used to inform the ranking statistic and improve GW signal identification.
The speed difference between the BAYESTAR and LALInference algorithms is mainly due to LALInference marginalizing over intrinsic parameters such as source mass, whereas these are fixed near the peak of the likelihood by BAYESTAR, which costs only a small loss of accuracy.
In this work we present a new method~\cite{url_sphradiometer} that reduces the localization time further by fixing additional parameters such as the distance to the source and orbit inclination.
This brings with it a yet further loss of accuracy, but provides an algorithm that fills a niche for ultra-fast location estimates.

Other source localization techniques, such as~\cite{rapid_pseudo_bayestar}, have been developed with a similar objective.
The work~\cite{rapid_pseudo_bayestar} is motivated by BAYESTAR~\cite{bayestar, bayestar3d}, which is the different approach from our new method built up on Excess power method.
Although the motivation of our new method is similar to the one of~\cite{rapid_pseudo_bayestar}, we compare performances between the new method and BAYESTAR, because BAYESTAR worked in the current detection pipeline.

\section{Notation}
We have GW detectors, LIGO-Hanford, LIGO-Livingston, Virgo, KAGRA~\cite{KAGRA1, KAGRA2} and so on.
Each detector outputs time series data.
From here, those are written as a vector:
\eq{
	\bm{d}[j] = \left( d_1[j], d_2[j], \cdots, d_D[j] \right)^T
}
where $j$ is an integer index enumerating discrete time, and $D$ is the number of detectors such that $D\geq 2$.
Fourier transformation to the $I$-th data whose length is $N$ is given by
\eq{
	\tilde{d}_I[k] &:= \sum_{j=0}^{N-1} d_I[j]\ \exp\left[ -2\pi\iu \frac{k}{N} j \right] \varDelta t
}
and
\eq{
	d_I[j] &:= \sum_{k=0}^{N-1} \tilde{d}_I[k]\ \exp\left[ 2\pi\iu \frac{k}{N} j \right] \varDelta f
}
where $\varDelta t$ and $\varDelta f$ are units of, respectively, discrete time and frequency satisfying $\varDelta t \varDelta f = 1/N$.
Also corresponding antenna responses~\cite{maggiore1} with polarization angle $\psi$ for GW source direction $\bm{\Omega}$ are written as a matrix:
\eq{
	\bm{F}(\bm{\Omega},\psi) = \left(
		\begin{array}{ccc}
			F_{1,+}(\bm{\Omega},\psi) & \cdots & F_{D,+}(\bm{\Omega},\psi) \\
			F_{1,\cross}(\bm{\Omega},\psi) & \cdots & F_{D,\cross}(\bm{\Omega},\psi) 
		\end{array}
		\right)^T
}

The data is composed of the signal and noise $d_I[j] = F_{I,+} h_\mathrm{GW,+}[j] + F_{I,\cross} h_\mathrm{GW,\cross}[j] + n_I[j]$, where $h_\mathrm{GW,+/\cross}$ is the GW for each of the two modes.
We assume that the noise $n_I[j]$ is uncorrelated between any pair of detectors:
\eq{
	\langle \tilde{n}_I[k] \tilde{n}^*_J[k'] \rangle := \frac{1}{2} \delta_{IJ} \frac{\delta_{kk'}}{\varDelta f} S_{n,I}[k]
}
where $S_{n,I}$ is noise power spectral density (PSD) for an $I$-th detector.
Using this, complex SNR is defined as follows,
\eq{
	\rho_I[j] &:= (d_I[j+j'] \,|\, h[j']) \\
	&:= 4 \sum_{k=0}^{N-1} \frac{\tilde{d}_I^*[k] \tilde{h}[k]}{S_{n,I}[k]} \exp\left[ 2\pi\iu \frac{j}{N}k \right] \varDelta f
}
where $h = h_+ + \iu h_\cross$ is a template, that is, a theoretical waveform normalized by $(h[j]|h[j]) = 2$.Then, SNR PSD is defined as noise PSD of the output of the matched filter (See \appref{SEC:PSD_SNR})
\eq{
	\left. \av{ \tilde{\rho}_I[k] \tilde{\rho}^*_J[k'] } \right|_{d=n} &= \frac{1}{2} \delta_{IJ} \frac{\delta_{kk'}}{\varDelta f} S_{\rho,I}[k] \\
	S_{\rho,I}[k] &:= 4\,\frac{\tilde{h}[k] \tilde{h}^*[k]}{S_{n,I}[k]} \label{EQ:S_rho}
}

Then, the data of $I$-th detector on (discrete) time domain is shifted to represent the data at geocenter.
If GW comes from $\bm{\Omega}$, the discrete time delay is $\tau_I(\bm{\Omega}) := \bm{r_I \cdot \Omega} / (c \varDelta t)$ between $I$-th detector at $\bm{r}_I$ and geocenter.
When this delay is applied, the time shifted data on frequency domain should be written as
\eq{
	\tilde{\bm{d}}[k;\bm{\Omega}] &= \left(
		\begin{array}{c}
			\tilde{T}_1[k; \bm{\Omega}] \tilde{d}_1[k]\\
			\vdots \\
			\tilde{T}_D[k; \bm{\Omega}] \tilde{d}_D[k]
		\end{array}
		\right)
}
where
\eq{
	\tilde{T}_I[k;\bm{\Omega}] &:= \exp\left[ 2\pi\iu \frac{k}{N} \tau_I(\bm{\Omega}) \right]
}
is the time delay operator.
For this data, time shifted SNR series are written as $\bm{\rho}[j;\bm{\Omega}] := (\bm{d}[j+j';\bm{\Omega}] | h[j'])$.

The objective is to compute the likelihood of obtaining a vector of SNR data given a waveform model which, here, includes the sky location.
Because they have been obtained from a colored linear filter, SNR time series samples are not independent random variables, therefore we obtain the likelihood using the standard frequency-domain whitening transformation, for which we introduce the whitened SNR and whitened time shifted SNR:
\eq{
	\hat{\tilde{\bm{\rho}}}[k] = \left(
		\begin{array}{c}
			\tilde{\rho}_1[k] / \sqrt{S_{\rho,1}[k]} \\
			\vdots \\
			\tilde{\rho}_D[k] / \sqrt{S_{\rho,D}[k]}
		\end{array}
		\right) \label{EQ:def_whitened_SNR}
}
\eq{
	\hatilde{\rho}_I[k;\bm{\Omega}] &= \tilde{T}_I[k;\bm{\Omega}] \hatilde{\rho}_I[k]
}
and whitened antenna response:
\eq{
	\bm{\hat{F}}[k; \bm{\Omega},\psi] :=& \left( \bm{\hat{F}_+}[k; \bm{\Omega}, \psi], \bm{\hat{F}_\cross}[k; \bm{\Omega}, \psi] \right) \\
		:=& \left(
		\begin{array}{c}
			F_{1,+}(\bm{\Omega},\psi)/S_{n,1}[k] \sqrt{S_{\rho,1}[k]},\\
			F_{1,\cross}(\bm{\Omega},\psi)/S_{n,1}[k] \sqrt{S_{\rho,1}[k]},
		\end{array}
		\right. \nonumber \\
		& \left.
		\begin{array}{cc}
			 \qquad \cdots, & F_{D,+}(\bm{\Omega},\psi)/S_{n,D}[k] \sqrt{S_{\rho,D}[k]} \\
			 \qquad \cdots, & F_{D,\cross}(\bm{\Omega},\psi)/S_{n,D}[k] \sqrt{S_{\rho,D}[k]}
		\end{array}
		\right)^T
}
This whitening SNR and antenna response simplify the formalism in the next section.

Under this notation, if GW is contained in data, a whitened SNR frequency series is written as
\eq{
	\hatilde{\rho}_I[k;\bm{\Omega}] =& \left( \hat{F}_{I,+}[k; \bm{\Omega}, \psi], \hat{F}_{I,\cross}[k; \bm{\Omega}, \psi] \right) \begin{pmatrix} \tilde{h}_+[k] \\ \tilde{h}_\cross[k] \end{pmatrix} \tilde{h}^*_{\mathrm{GW}}[k] \nonumber\\
		&+ \frac{\tilde{n}^*_I[k] (\tilde{h}_+[k] + \tilde{h}_\cross[k])}{S_{n,I}[k]} \frac{\exp\left[ 2\pi\iu\frac{k}{N} \tau_I(\bm{\Omega}) \right]}{\sqrt{S_{\rho,I}[k]}} \label{EQ:whitened_SNR}
}
where $h_\mathrm{GW} = h_\mathrm{GW,+} + h_\mathrm{GW,\cross}$, which is introduced for a simple formalization.
A detail of the derivation is in \appref{SEC:whitened_SNR}.
We will use this series for the localization instead of the strain data.

\section{Compact Binary Coalescence parametrized Likelihood} \label{SEC:CBC_parameterized_Likelihood}
Here, we assume that the source is CBC without precession, therefore the two polarizations are related by $\tilde{h}_\cross = \iu\beta\tilde{h}_+$ where $\beta = \frac{2\cos\iota}{1+\cos^2\iota}$ with the inclination $\iota$.
Then if the noise is Gaussian, the probability of obtaining $\bm{\hatilde{\rho}}$ in the presence of $\tilde{h}_{\mathrm{GW}}$ with given parameters $\bm{\Omega}, \beta, \psi$ is
\eq{
	&p(\bm{\hatilde{\rho}}|\bm{\Omega}, \tilde{h}_{\mathrm{GW}}, \beta, \psi) \nonumber\\
	\propto& \exp\left[ -2 \sum_{k=0}^{N-1} \left| \bm{\hatilde{\rho}}[k; \bm{\Omega}] \right.\right.\nonumber\\
	& \quad \left.\left. - \left( \bm{\hat{F}_+}[k; \bm{\Omega}, \psi] + \iu\beta \bm{\hat{F}_\cross}[k; \bm{\Omega}, \psi] \right) \tilde{h}_+[k] \tilde{h}_{\mathrm{GW}}[k]  \right|^2 \varDelta f \right]
}
Since $\tilde{h}_{\mathrm{GW}}$ is not known a priori, this probability should be maximized with respect to $\tilde{h}_{\mathrm{GW}}$.
This was solved by Sutton et al. in~\cite{Sutton} for the case of general GWs.
This probability is maximized by $\tilde{h}_+ \tilde{h}_{\mathrm{GW}} = \left[ \left| \bm{\hat{F}_+} \right|^2 + \beta^2 \left| \bm{\hat{F}_\cross} \right|^2\right]^{-1} \left( \bm{\hat{F}_+} + \iu\beta \bm{\hat{F}_\cross} \right)^\dagger \bm{\hatilde{\rho}}$, which is in effect maximizing the probability over the distance to the source for arbitrary choice of inclination parameter $\beta$ and polarization angle $\psi$
\eq{
	p(\bm{\hatilde{\rho}}|\bm{\Omega}, \beta, \psi) &\propto \exp\left[ 2 \sum_{k=0}^{N-1} \bm{\hatilde{\rho}}^\dagger[k;\bm{\Omega}] \bm{\hat{P}}[k; \bm{\Omega}, \beta, \psi] \bm{\hatilde{\rho}}[k;\bm{\Omega}] \right] \label{EQ:bare_probability}
}
where
\begin{widetext}
	\eq{
		\bm{\hat{P}}[k; \bm{\Omega}, \beta, \psi] &:= \frac{(\bm{\hat{F}}_+[k; \bm{\Omega}, \psi] + \iu\beta \bm{\hat{F}}_\cross[k; \bm{\Omega}, \psi]) \otimes (\bm{\hat{F}}_+[k; \bm{\Omega}, \psi] - \iu\beta \bm{\hat{F}}_\cross[k; \bm{\Omega}, \psi])}{ \left| \bm{\hat{F}_+}[k; \bm{\Omega}, \psi] \right|^2 + \beta^2 \left| \bm{\hat{F}_\cross}[k; \bm{\Omega}, \psi] \right|^2  } \label{EQ:def_P}
	}
\end{widetext}
is a projection operator to extract the GW component from the data\footnote{$\bm{\hat{P}}$ satisfies $\bm{\hat{P}} \bm{\hat{P}} = \bm{\hat{P}}$ and $\bm{\hat{P}} (\bm{\hat{F}_+} + \iu\beta \bm{\hat{F}_\cross}) \tilde{h}_+ \tilde{h}_{\mathrm{GW}} = (\bm{\hat{F}_+} + \iu\beta \bm{\hat{F}_\cross}) \tilde{h}_+ \tilde{h}_{\mathrm{GW}}$.
Therefore $\bm{\hat{P}}$ is a projection operator to extract GW contributions from the given data.
Since $\bm{\hat{P}}$ is constructed by only $\bm{\hat{F}_+} + \iu\beta \bm{\hat{F}_\cross}$, the dimension of the GW space where $\bm{\hat{P}}$ project data onto is one (see \figref{FIG:projection}).}. Also, $\otimes$ is direct product.
\myfig{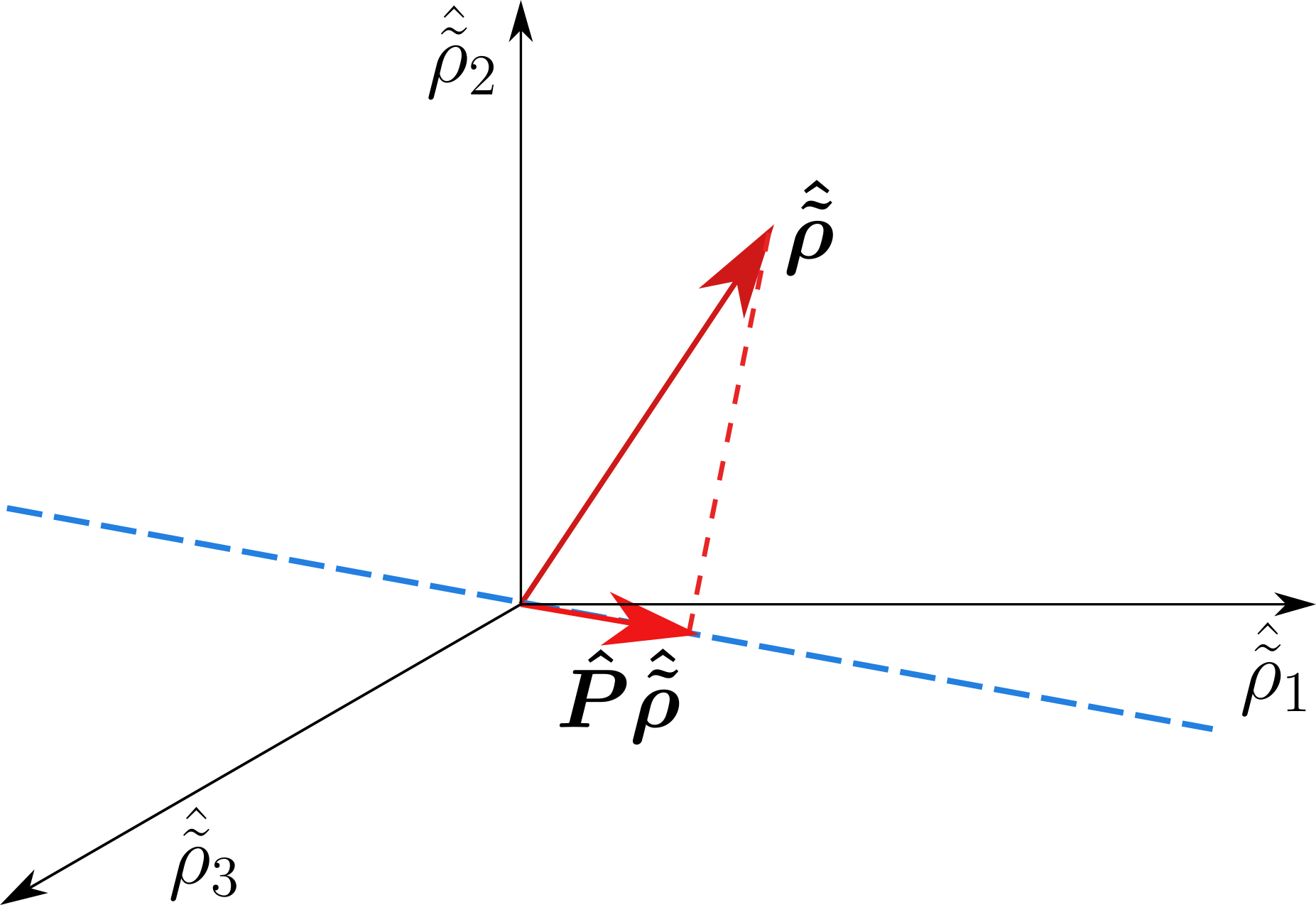}{Schematic representation of the concept of $\bm{\hat{P}}$ for a three detectors case, which is in a SNR data space spanned by those detectors.  Red vector $\bm{\hatilde{\rho}}$ is an observed data in the data space.  Blue dashed line is the GW space.}{FIG:projection}

We now maximize \eqref{EQ:bare_probability} with respect to the remaining two parameters $\beta, \psi$.
The solutions of $0 = \frac{\partial}{\partial \beta} p(\bm{\hatilde{\rho}}|\bm{\Omega}, \beta, \psi) = \frac{\partial}{\partial \psi} p(\bm{\hatilde{\rho}}|\bm{\Omega}, \beta, \psi)$ are $\beta = 0, \pm 1$ with $\psi$ depending on SNR data $\bm{\hatilde{\rho}}$ respectively, which are extremizing points.
We cannot estimate which condition is a global maximum before observations, therefore we simply marginalize \eqref{EQ:bare_probability} over the extremizing points, $\beta = 0, \pm1$.
Here, remind that our localization target is a detected CBC.
However the likelihood $p(\bm{\rho} | \bm{\Omega}, \beta, \psi)$ does not include the condition and then a prior should be needed.
In this condition, from \appref{SEC:prior_of_beta} the marginalized posterior for the detected GW is derived with a prior of $\beta$ for the detected GWs, $p(\beta|\mathrm{detect}) \sim (\delta_{\beta=+1} + \delta_{\beta=-1}) / 2$, which drops the $\beta = 0$ term:
\eq{
	p(\bm{\hatilde{\rho}}|\bm{\Omega}, \psi) \propto& \sum_{\beta=0,\pm 1} p(\bm{\rho}|\bm{\Omega}, \beta, \psi) p(\beta|\mathrm{detect})\\
		\propto& \sum_{\beta=\pm1} \exp\Biggl[ 2 \sum_{IJ\in IFO} \sum_{k=0}^{N-1} \hatilde{\rho}_I^*[k] \hatilde{\rho}_J[k] \nonumber\\
		& \quad \times \hat{P}_{IJ}[k;\bm{\Omega}, \beta, \psi=0] \tilde{T}_I^*[k;\bm{\Omega}] \tilde{T}_J[k;\bm{\Omega}] \Biggr] \label{EQ:basic_probability}
}
where the projection operator $\bm{\hat{P}}$ for $\beta=\pm 1$ can neglect a dependence of $\psi$, so that $\psi = 0$ is set\footnote{$\bm{\hat{F}_+} \pm \iu\bm{\hat{F}_\cross} \rightarrow \left( \bm{\hat{F}_+} \pm \iu\bm{\hat{F}_\cross} \right) \exp\left( \pm \iu\psi \right)$ is satisfied by rotating $\psi$; $\bm{\hat{F}_+} \rightarrow \bm{\hat{F}_+} \cos2\psi + \bm{\hat{F}_\cross} \sin2\psi$, $\bm{\hat{F}_\cross} \rightarrow -\bm{\hat{F}_+} \sin2\psi + \bm{\hat{F}_\cross} \cos2\psi$.
This phase factor is canceled in $\bm{\hat{P}}$}.

Coalescence phase has not been fixed yet because of maximizing $\tilde{h}_\mathrm{GW} \propto \hatilde{\rho}$, which is embedded in the complex phase of $\rho_I$ at the merger, and is same value for all $\rho_I$.
Then, the coalescence phase dependence can be neglected because \eqref{EQ:basic_probability} does not depend on it.

Here, $\hat{P}_{IJ} \tilde{T}^*_I \tilde{T}_J$ is independent of the SNR data, allowing it to be pre-computed for speed.
Following the approach presented in~\cite{Kipp}, we expand the $\rho$-independent factor in spherical harmonics $Y_{lm}$:
\eq{
	p(\bm{\hatilde{\rho}}|\bm{\Omega}) \propto& \sum_{\beta=\pm1} \exp\left[ 2 \Re \sum_{lm} \left\{ \sum_{IJ\in \mathrm{IFO}} \sum_{k=0}^{N-1} \right.\right. \nonumber\\
		& \left.\left. \left( \hat{P} \tilde{T}^* \tilde{T} \right)_{IJ}^{lm}[k; \beta] \hatilde{\rho}_I^*[k] \hatilde{\rho}_J[k] \right\} Y_{lm}(\bm{\Omega}) \right]
}
To calculate this inverse spherical harmonic transformation, healpy\footnote{http://healpix.sf.net}~\cite{healpy, healpix} is used.

In this paper, an isotropic prior is assumed to obtain a posterior:
\eq{
	p(\delta, \alpha) = \frac{1}{4\pi} \cos\delta \label{EQ:declination_prior}
}
where $\delta$ is declination and $\alpha$ is right ascension.

Hence one gets a below posterior from Bayes' theorem:
\eq{
	p(\bm{\Omega}|\bm{\rho}) \propto p(\bm{\rho}|\bm{\Omega}) p(\delta, \alpha) \label{EQ:sphrad_posterior}
}
We use this probability to produce sky maps.

\section{Regulator}
In the definition of the whitened SNR in \eqref{EQ:def_whitened_SNR} the ratio $\tilde{\rho}[k] / \sqrt{S_\rho[k]}$ is not well defined for all frequency bins $k$.
In particular, because inspiral templates have $0$ signal energy above some high-frequency cutoff $S_\rho$ is $0$ for some $k$ and the whitened SNR is undefined.
In future work, this problem will be addressed with a more sophisticated treatment, but at present we have found it is sufficient to regulate the instability by multiplying each term in $p(\bm{\hatilde{\rho}}|\bm{\Omega})$ by $2\sqrt{S_{\rho I}[k]} \sqrt{S_{\rho J}[k]}$\footnote{SNR PSD is defined as the double-sided PSD.
However, on discrete domain, the single-sided PSD is used.
Then, the factor $2$ is needed.}.
By this process, the whitened SNR series become SNR series, and then especially auto-correlation terms differ from an expected values.
Thus, to obtain a reasonable value, a probability from cross-correlation terms $p(\hatilde{\bm{\rho}} | \bm{\Omega})$ is used for the localization.
Also, by the same unstable reason, $\hat{\bm{P}}$ is replaced with $\bm{P}(\bm{\Omega}, \beta, \psi=0) := \left. \hat{\bm{P}} \right|_{\hat{\bm{F}} = \bm{F}}$\footnote{$\bm{\hat{P}} \rightarrow \bm{P}$ corresponds to an assumption that all detector have same PSD because, if so, $\sqrt{S_\rho}$ in denominator and numerator are canceled.}.:
\eq{
	p^{\mathrm{cross}}(\bm{\hatilde{\rho}}|\bm{\Omega}) \propto& \sum_{\beta=\pm1} \exp\left[ 8 \Re \sum_{lm} \left\{ \sum_{I>J\in \mathrm{IFO}} \sum_{k=0}^{N-1} \right.\right. \nonumber\\
		& \left.\left. \left( P \tilde{T}^* \tilde{T} \right)_{IJ}^{lm}[k; \beta] \tilde{\rho}_I^*[k] \tilde{\rho}_J[k] \right\} Y_{lm}(\bm{\Omega}) \right]
}

%

\section{Results and Discussion} \label{SEC:results}
We will compare the new method with current methods, BAYESTAR~\cite{bayestar, bayestar3d}.

\subsection{Injection test}
\begin{figure}[t]
	\centering
	\begin{tabular}{c}
		\begin{minipage}{0.5\linewidth}
			\centering
			\includegraphics[width=\linewidth]{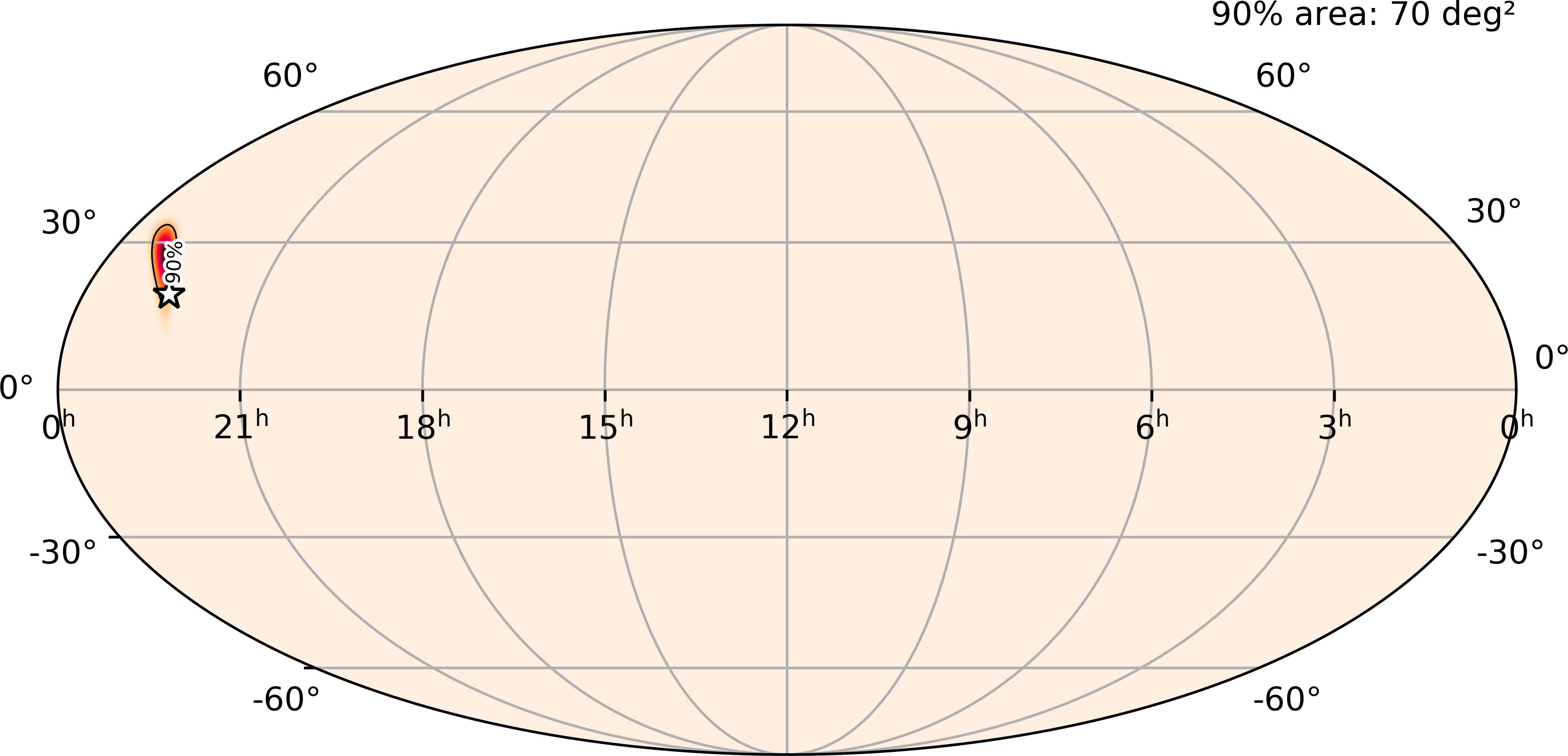}\\
			\includegraphics[width=\linewidth]{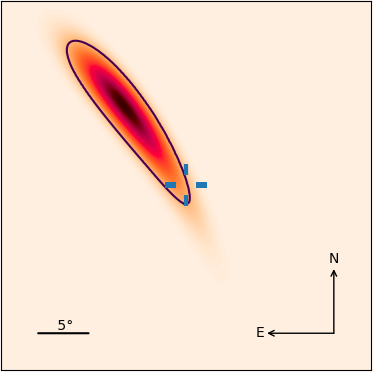}\\
			New method
		\end{minipage}

		\begin{minipage}{0.5\linewidth}
			\centering
			\includegraphics[width=\linewidth]{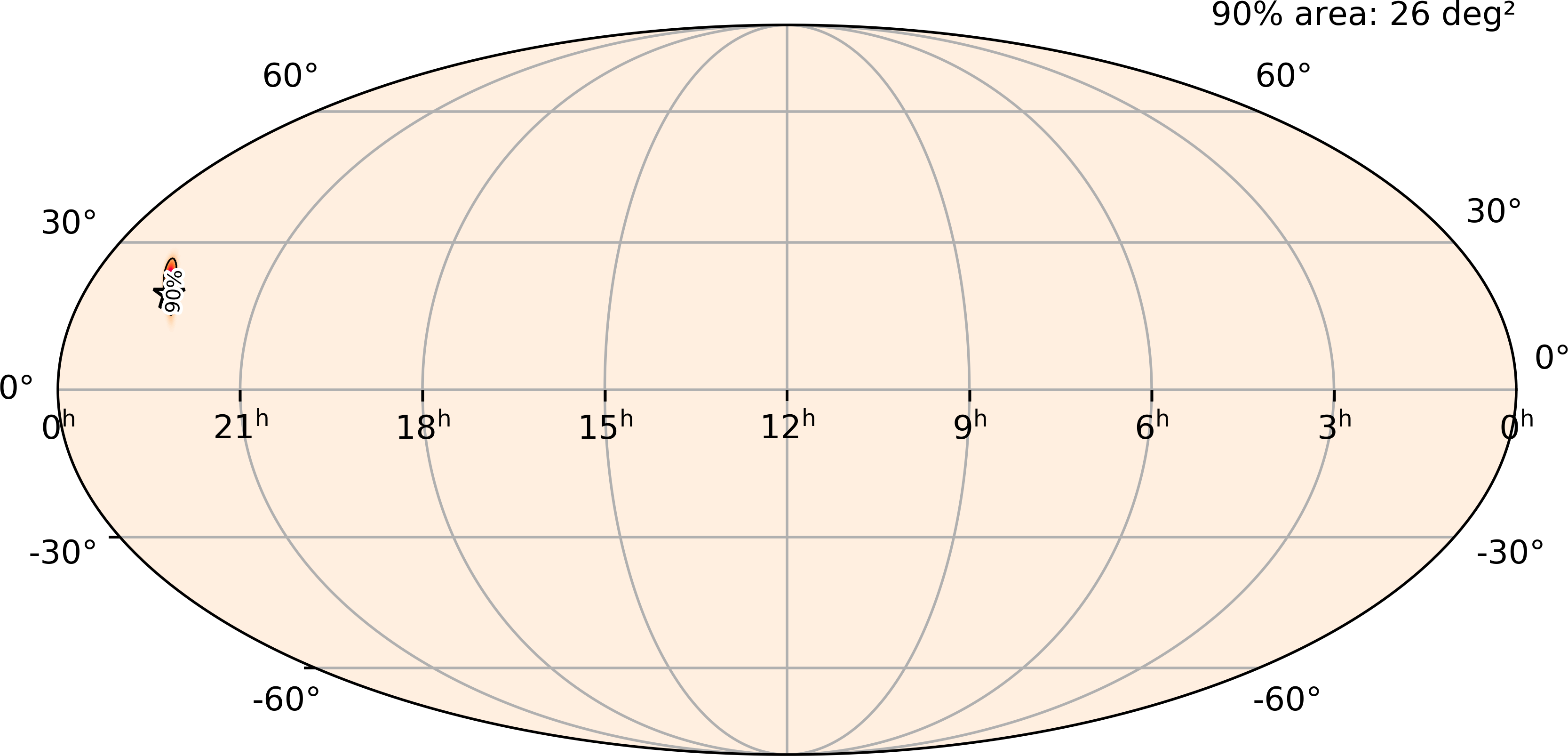}\\
			\includegraphics[width=\linewidth]{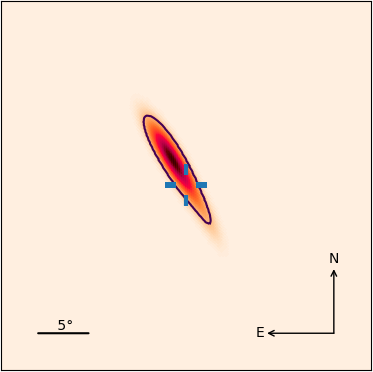}\\
			BAYESTAR
		\end{minipage}
	\end{tabular}
	\caption{All sky and zoom maps of the localization results of the new method and BAYESTAR~\cite{bayestar, bayestar3d} for one of the injections with $(\mathrm{SNR}_{\mathrm{Hanford}}, \mathrm{SNR}_{\mathrm{Livingston}}, \mathrm{SNR}_{\mathrm{Virgo}}) = (22.0, 15.8, 7.24)$, $m_1 = 1.76 M_\odot, m_2 = 1.06 M_\odot$ and no spin at $\SI{1187094629.9577243}{s}$ at geocenter.  The right ascension and the declination are respectively $\SI{22.5}{hour}$ and $19.2^\circ$ marked by a star in all sky maps and a blue plus in zoom maps.  Purple line is a $90\%$ contour whose region size is $\SI{70}{deg^2}$ for the new method and $\SI{26}{deg^2}$ for BAYESTAR.  In this example, both methods have the true direction inside the $90\%$ contour.}
	\label{FIG:injection_example}
\end{figure}

We evaluated the performance from an injection test.
The setup is below:
\begin{itemize}
	\item TaylorT4threePointFivePN was injected into second observing run (O2) data from $\SI{1186624818}{s}$ to $\SI{1187312718}{s}$ in GPS time, that is, August 13-21 in 2017.
	\item The three detectors, LIGO-Hanford, LIGO-Livingston and Virgo were used.
	\item The component masses are randomly sampled for $m_1, m_2 \in [1.08 M_\odot, 1.58 M_\odot]$ from Gaussian with the mean of $1.33 M_\odot$ and the standard deviation of $0.05 M_\odot$.
	\item no component spins.
	\item The distance was randomly sampled from a log-uniform distribution for $r \in [\SI{20}{Mpc}, \SI{200}{Mpc}]$.
	\item $(\alpha, \delta)$ and $(\iota, \psi)$ were distributed isotropically.
	\item Triggers which is searched with the matched filter technique of a CBC detection software, GstLAL~\cite{gstlal1, gstlal2} were selected with satisfying:
	\begin{itemize}
		\item Those are contained within $\SI{1}{s}$ around injected time.
		\item The SNRs of more than two detectors are exceeded over $8$.
		\item All detectors are worked on Science mode.
		\item The network SNR $\sqrt{\sum_{I\in \mathrm{IFO}} \mathrm{SNR}^2_I}$ is maximized in the triggers.
	\end{itemize}
	\item $935$ injections were used.
\end{itemize}
Under the above setting, complex SNR time series are generated in $\SI{0.17}{s}$ around the triggered time when detecting the trigger.
\figref{FIG:injection_example} is an example of the localization of the injections.

\subsubsection{Consistency}
From the above complex SNR time series, We produce skymaps and a p-p plot (\figref{FIG:pp-plot}) for the new method and BAYESTAR~\cite{bayestar, bayestar3d}.
From the definition of $p$ value, the fraction of the injections with a $p$ from the peak of maps to the injected direction should be equal to the $p$, that is, the cumulative lines should be on the diagonal.
From \figref{FIG:pp-plot}, the average of the cumulative line of the new method is on the diagonal.
Then, the average of the new method is statistically consistent.
Nevertheless, parts of the cumulative line are out of the $95\%$ error region.
The origin should be from the approximation of $\bm{\hat{P}} \rightarrow \bm{P}$ (see \secref{SEC:CBC_parameterized_Likelihood}), because both methods assumed Gaussian noise and CBC waveform, that is, the difference was from the other.
That approximation is the sole one to be able to shift the peak of maps.
\myfig{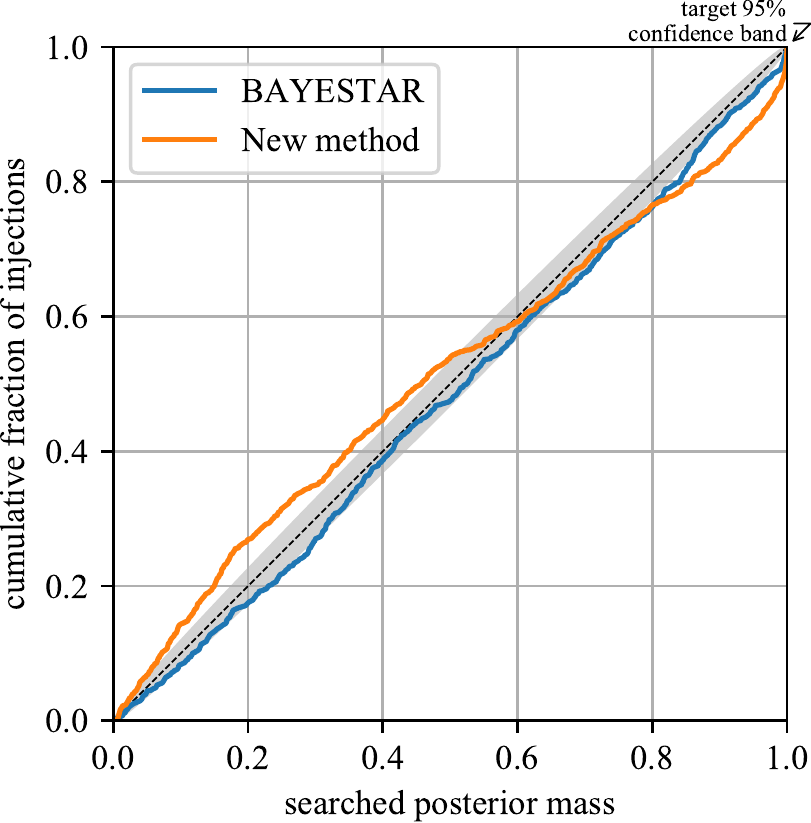}{p-p plot\cite{url_ligo_skymap} of the new method and BAYESTAR.  Cumulative fractions of the injections are a ratio included in a $p$ value.  Gray region is error region in $95\%$.}{FIG:pp-plot}

\subsubsection{Accuracy}
\begin{table}[tb]
	\centering
	\caption{Cherenkov Telescope Array has three size telescopes, SST, MST and LST~\cite{CTA}.}
	\begin{tabular}{cccc}
		Name & Field of view & Target energy & Slew speed \\ \hline
		SST & $\SI{8.8}{deg}$ & $1-300\unit{TeV}$ & $\lesssim \SI{1}{min}$ \\
		MST & $7.5-7.7\unit{deg}$ & $\SI{80}{GeV}-\SI{50}{TeV}$ & $< \SI{90}{s}$ \\
		LST & $\SI{4.5}{deg}$ & $\SI{20}{GeV}-\SI{3}{TeV}$ & $< \SI{20}{s}$ \\
	\end{tabular}
	\label{TAB:FOVs}
\end{table}

From the used detectors, the maximum spherical harmonic degree $l$ is $184$.
Then the accuracy is relevant by $\sim 1^\circ$.

\figref{FIG:error_size} is the area size distribution recognized as accuracy.
Then, the square root of it can be recognized as the opening angle which the telescopes require.
From \figref{FIG:error_size}, the new method is about $10$ times less accurate than BAYESTAR~\cite{bayestar, bayestar3d}.
Since the area size is $\sim \SI{65}{deg}^2$, the opening angle is $\sim \SI{8}{deg}$.
This opening angle is comparable with the field of view of the cherenkov telescope array (CTA) (see \tabref{TAB:FOVs}), so that it is sufficiently accurate for early warning.
This worse accuracy than BAYESTAR should be due to the regulator, that is, no whitening approximation (see \secref{SEC:CBC_parameterized_Likelihood}).
However, the no whitening effect should be recognized as $3.7$ times rather $10$ times, from \figref{FIG:error_size_ratio} which compares those area sizes for each injections.
Summarizing the above, the new method and BAYESTAR have complementary relation with each other in terms of speed and accuracy.
Using more information by the marginalizing, BAYESTAR is more robust than the new method.
Therefore BAYESTAR should have better accuracy even if all approximations are removed.
\myfig{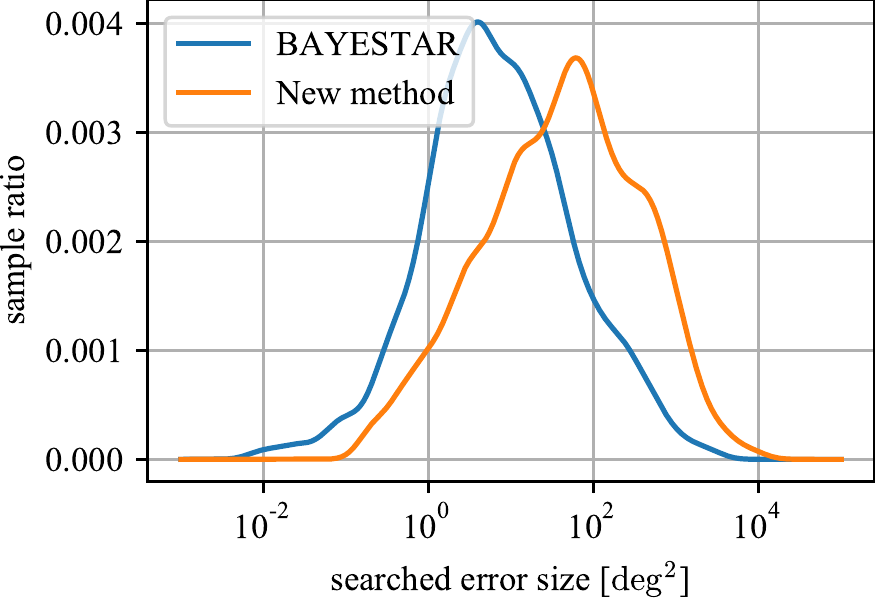}{Area size distribution of pixels from the peak of maps to the injected direction. Sample ratio is a ratio with an area size.}{FIG:error_size}
\myfig{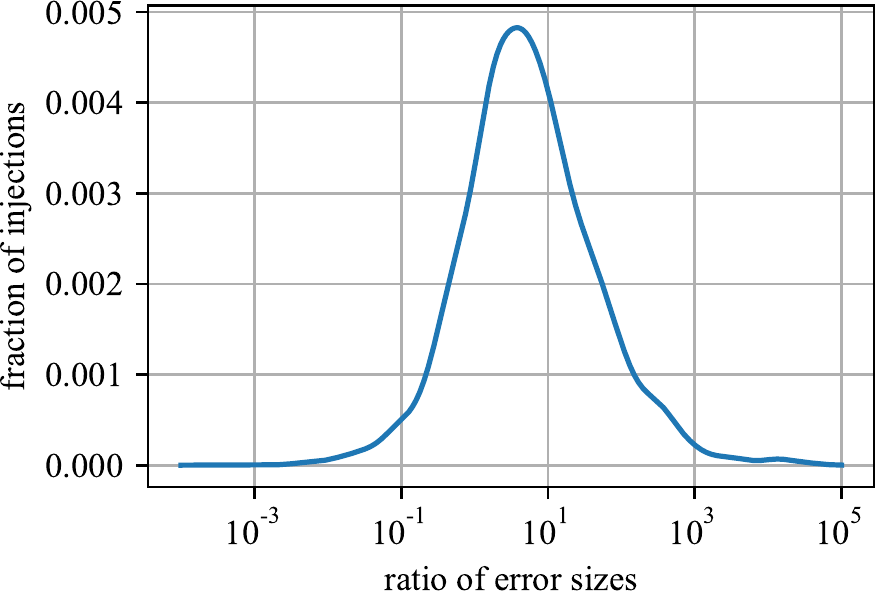}{Area size ratio distribution between the new method and BAYESTAR for each injections.  The ratio is (area of the new method) / (area of BAYESTAR).}{FIG:error_size_ratio}

\subsubsection{Computational Cost}
The main advantage of the new method is its reduced computational cost and its speed.
We measured the relative computational cost of this algorithm and BAYESTAR in single-threaded mode on an Intel Core i7-7600U CPU @ $\SI{2.80}{GHz}$, and also measured the relative run-times of BAYESTAR in that configuration to a fully parallel configuration on an Intel Xeon Gold 6136 CPU @ $\SI{3.00}{GHz}$.
Taking the single-threaded run times to be dominated by arithmetic operations (I/O is not significant) then this comparison provides an estimate of the ratio of arithmetic operation count required by the two techniques to produce a location estimate.
BAYESTAR is a mature code that has been optimized for the highly parallel Xeon hardware, so we also report a speed comparison of the BAYESTAR code in its production configuration.
\begin{center}
	\centering
	\begin{tabular}{c|cc}
			& New method & BAYESTAR \\ \hline
		single-threaded & $\SI{0.73}{s}$ & $\SI{47}{s}$ \\
		parallelized & - & $\SI{3.3}{s}$
	\end{tabular}
\end{center}

\section{Summary and Future work} \label{SEC:summary}
We developed a rapid localization method which is $64$ times faster than BAYESTAR~\cite{bayestar, bayestar3d} at the cost of accuracy by an order of magnitude.

Our method assumes the Gaussian noise.
To estimate the direction, the new method takes into account the time delays, the amplitude ratios and the phase differences between SNR time series from different detectors.
By maximizing or marginalizing the probability model \eqref{EQ:bare_probability} and extracting precalculated factors, the number of parameters to estimate during the calculation is reduced, which leads to speeding up the localization.

The new method has three differences from Excess power method~\cite{Sutton} and BAYESTAR~\cite{bayestar, bayestar3d} as follows:
\begin{enumerate}
	\item Compared to BAYESTAR which marginalizes the posterior sky map over distance to source and source orbit inclination, the new method maximizes the posterior with respect to these two parameters.  This sacrifices some accuracy in the map, but allows for some expressions to be factored into terms that depend only on data and terms that do not, which can then be pre-computed for greater speed.
	\item SNR time series are used instead of strain data.  By this, one can generate sky maps optimized for CBC templates, and suppress the noise contamination which is orthogonal to the template.  This is the difference from Excess power method.
	\item The CBC parametrization is used instead of the general parametrization used by Excess power method.  By this, our target is only CBC, which is same as BAYESTAR.  Then, the new method is more accurate than Excess power method, but not BAYESTAR.  Also the new method can localize GW sources for more than single detector working case but Excess power method can localize for more than the double detector working case.
\end{enumerate}

As a potential of further improvements, the approximations applied in \secref{SEC:CBC_parameterized_Likelihood} are enumerated:
\begin{enumerate}
	\item All detectors have the same PSD, that is, neglecting frequency dependence of Projection operator to correct distortions from the antenna responses and extract the GW components from data: $\bm{\hat{P}}[k;\bm{\Omega},\beta,\psi=0] \rightarrow \bm{P}(\bm{\Omega},\beta,\psi=0)$.
	\item The PSDs of SNR time series are flat, that is, no whitening approximation: $\hatilde{\rho}[k] \rightarrow \tilde{\rho}[k]$.  Since, by this regulator, the auto-correlation terms become far from expected values, those terms are neglected.
\end{enumerate}
The both approximations are meant to avoid numerical instability.
Removing these approximations is future work.
First one could shift the peak of maps to the correct peak because our probability should be more affected from the detector with higher sensitivities (more likely).
Second one could make error region of sky maps wavy (smaller) because it makes complex phase variation fast, and our probability picks up just real part from the correlations.

\begin{acknowledgments}
This research has made use of data, software and/or web tools obtained from the Gravitational Wave Open Science Center (https://www.gw-openscience.org), a service of LIGO Laboratory, the LIGO Scientific Collaboration and the Virgo Collaboration.
LIGO is funded by the U.S. National Science Foundation.
Virgo is funded by the French Centre National de Recherche Scientifique (CNRS), the Italian Istituto Nazionale della Fisica Nucleare (INFN) and the Dutch Nikhef, with contributions by Polish and Hungarian institutes.
This work was supported by the International Graduate Program for Excellence in Earth-Space Science (IGPEES).
We would like to thank Heather Fong, Duncan Meacher, Cody Messick and Leo Pound Singer for teaching us how to use the analyzing software.
Also, we are grateful for LIGO-Virgo's computational resources and the data~\cite{open_data_appreciate}, because the injection data sets are produced with those.
Some of the results in this paper have been derived using the healpy and HEALPix packages.
\end{acknowledgments}

\appendix
\section{Derivation of the Power Spectral Density of Signal to Noise Ratio} \label{SEC:PSD_SNR}
We derive the PSD of SNR with no GW.
For uncorrelated noise,
\eq{
	& \frac{1}{2} \delta_{I J} \delta(f-f') S_{\rho I}(f) \\
		=& \langle \tilde{\rho}_I(f) \tilde{\rho}^*_J(f') \rangle \\
		=& \int \diff t \diff t' \langle \rho_I(t) \rho_J^*(t') \rangle \e^{-2\pi\iu (ft - f't')} \\
		=& 4 \int \diff t \diff t' \int_{-\infty}^\infty \diff g \diff g' \frac{\langle \tilde{n}_I(g) \tilde{n}_J^*(g') \rangle \tilde{h}(g) \tilde{h}^*(g')}{S_{n,I}(g) S_{nJ}(g')} \nonumber \\
		 & \quad \times \e^{-2\pi\iu (ft - f't') + 2\pi\iu (gt - g't')} \\
		=& \frac{\delta_{IJ}}{2} 4 \int_{-\infty}^\infty \diff g \frac{\tilde{h}(g) \tilde{h}^*(g)}{S_{n,I}(g)} \delta(f-g) \delta(f'-g) \\
		=& \frac{1}{2} \delta_{IJ} \delta(f-f') \, 4 \, \frac{\tilde{h}(f) \tilde{h}^*(f)}{S_{n,I(f)}}
}
Comparing LHS with RHS, \eqref{EQ:S_rho} is obtained.

\section{Derivation of \eqref{EQ:whitened_SNR}} \label{SEC:whitened_SNR}
In nature, true GW on time domain should be written as $h_\mathrm{GW}[j] = h_{\mathrm{GW}, +}[j] + h_{\mathrm{GW}, \cross}[j] \in \mathbb{R}$.
For example, $h_{\mathrm{GW}, +}$ is $\cos$-mode and $h_{\mathrm{GW}, \cross}$ is $\sin$-mode.
The output data with the GW contaminated by noise $n_I$ from $I$-th detector is $d_I[j] = F_{I,+} h_{\mathrm{GW}, +}[j] + F_{I,\cross} h_{\mathrm{GW}, \cross}[j] + n_I[j]$.
To this data, the SNR time series for a complex template $h_+[j] + h_\cross[j]$ is
\eq{
	\rho_I[j] &= F_{I,+} (h_{\mathrm{GW},+}[j+j'] \,|\, h_+[j']) \nonumber \\
		&+ F_{I,\cross} (h_{\mathrm{GW}, \cross}[j+j'] \,|\, h_\cross[j']) \nonumber \\
		&+ (n_I[j+j'] \,|\, h_+[j'] + h_\cross[j'])
}
Assuming $(h_+ \,|\, h_{\mathrm{GW}, \cross}) = (h_\cross \,|\, h_{\mathrm{GW}, +}) = 0$,
\eq{
	\rho_I[j] &= F_{I,+} (h_{\mathrm{GW}}[j+j'] \,|\, h_+[j']) \nonumber \\
		&+ F_{I,\cross} (h_{\mathrm{GW}}[j+j'] \,|\, h_\cross[j']) \nonumber \\
		&+ (n_I[j+j'] \,|\, h_+[j'] + h_\cross[j'])
}
Doing Fourier transformation and whitening,
\eq{
	\hatilde{\rho}_I[k] &= \hat{F}_{I,+}[k; \bm{\Omega}, \psi] \tilde{h}_+[k] \tilde{h}^*_\mathrm{GW}[k] \tilde{T}^*_I[k; \bm{\Omega}] \nonumber \\
		&+ \hat{F}_{I,\cross}[k; \bm{\Omega}, \psi] \tilde{h}_\cross[k] \tilde{h}^*_\mathrm{GW}[k] \tilde{T}^*_I[k; \bm{\Omega}] \nonumber \\
		&+ \frac{\tilde{n}^*_I[k] (\tilde{h}_+[k] + \tilde{h}_\cross[k])}{S_{n,I}[k]} \frac{1}{\sqrt{S_{\rho,I}[k]}}
}
For convenience, time at geocenter is used.
Thus, shifting time $\tau_I(\bm{\Omega})$, we can obtain \eqref{EQ:whitened_SNR}.

\section{Prior of \texorpdfstring{$\beta$}{\textbackslash beta}} \label{SEC:prior_of_beta}
The probability of detecting GWs with an inclination $\iota$ should be proportional to an observable volume if the number density of CBC is uniform.
\eq{
	p(\mathrm{detect}|\iota) &\propto D_{\mathrm{range}}^3(\iota) \propto g^3(\iota)\\
	g(\iota) &:= \left( \frac{ 1 + \cos^2\iota }{2} \right)^2 + \cos^2\iota
}
where $D_\mathrm{range}$ is the range of detectors~\cite{Jolien, maggiore1}.
Since our universe should not have special direction, the prior is similar to \eqref{EQ:declination_prior}:
\eq{
	p(\iota) \propto \sin\iota
}
From Bayes' theorem, the probability of the inclination $\iota$ for GWs to be detected is
\eq{
	p(\iota|\mathrm{detect}) \propto p(\mathrm{detect}|\iota) p(\iota) \propto g^3(\iota) \sin\iota
}

Next, a probability of obtaining $\beta = 2\cos\iota / (1+\cos^2\iota)$ for detected events in general is derived.
From $|p(\iota|\mathrm{detect})\,d\iota| = |p(\beta|\mathrm{detect})\,d\beta|$,
\eq{
	p(\beta|\mathrm{detect}) &= p(\iota|\mathrm{detect}) \left| \frac{d\iota(\beta)}{d\beta} \right|\\
		&\propto g^3(\iota(\beta)) \left| \frac{\sin\iota(\beta)}{\beta(1-\beta^2)} \right| \sqrt{\sqrt{1-\beta^2} - (1-\beta^2)}\\
	\iota(\beta) &= \cos^{-1}\left[ \frac{1-\sqrt{1-\beta^2}}{\beta} \right]
}
\myfig{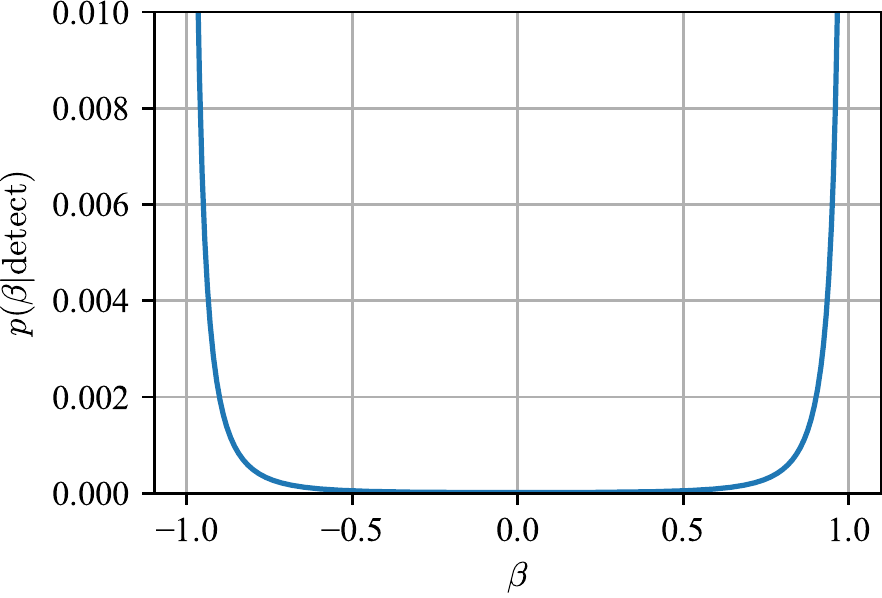}{This is a probability of $\beta$ for detected events, that is $\beta$ vs. $p(\beta|\mathrm{detect})$.}{FIG:prior_of_beta}
\figref{FIG:prior_of_beta} shows $p(\beta|\mathrm{detect})$ as a function of $\beta$. We note that $p(\iota|\mathrm{detect})\,d\iota$ is well-defined, that is,
\eq{
	\infty &> \int_0^\pi p(\iota|\mathrm{detect}) \,\diff \iota\\
		&= \int_{-1}^1 p(\beta|\mathrm{detect}) \,\diff \beta
}
Therefore $p(\beta|\mathrm{detect})$ is normalized.
Nevertheless, $p(\beta|\mathrm{detect})$ has strong peak for $\beta = \pm 1$.
Hence it is approximated with Kronecker-$\delta$:
\eq{
	p(\beta|\mathrm{detect}) \sim \frac{\delta_{\beta=+1} + \delta_{\beta=-1}}{2}
}
Considered in \secref{SEC:CBC_parameterized_Likelihood}, the probability model is extremized at $\beta = 0$ or $\pm1$.
Our purpose is not marginalizing posterior with respect to all $\beta$ but the extrema.
Therefore, this approximation is reasonable.

Despite of $\beta=\pm1$ ($\iota=0,\pi$), this prior does not mean that the detected GW is from the angle.
Statistically, the angle is just preferred from the fact that GW flux is highest along the orbital axes.


\bibliographystyle{unsrt}
\bibliography{references}

\begin{thebibliography}{10}

\bibitem{LIGO1}
J~Aasi, B~P Abbott, R~Abbott, T~Abbott, M~R Abernathy, K~Ackley, C~Adams,
  T~Adams, P~Addesso, et~al.
\newblock {Advanced LIGO}.
\newblock {\em Classical and Quantum Gravity}, 32(7):074001, Mar 2015.

\bibitem{LIGO2}
Gregory~M Harry.
\newblock {Advanced {LIGO}: the next generation of gravitational wave
  detectors}.
\newblock {\em Classical and Quantum Gravity}, 27(8):084006, apr 2010.

\bibitem{Virgo}
F~Acernese, M~Agathos, et~al.
\newblock {Advanced Virgo: a second-generation interferometric gravitational
  wave detector}.
\newblock {\em Classical and Quantum Gravity}, 32(2):024001, Dec 2014.

\bibitem{GW170817_observation}
B. P. Abbott, R.~Abbott, T. D. Abbott, F.~Acernese, K.~Ackley, C.~Adams,
  T.~Adams, P.~Addesso, R. X. Adhikari, V. B. Adya, et~al.
\newblock {GW170817: Observation of Gravitational Waves from a Binary Neutron
  Star Inspiral}.
\newblock {\em Physical Review Letters}, 119(16), Oct 2017.

\bibitem{GW170817_multimessenger}
B.~P. Abbott et~al.
\newblock {Multi-messenger Observations of a Binary Neutron Star Merger}.
\newblock {\em Astrophys. J.}, 848(2):L12, 2017.

\bibitem{CBC-GRB}
William~H. Lee, Enrico Ramirez-Ruiz, and Jonathan Granot.
\newblock {A Compact Binary Merger Model for the Short, Hard {GRB} 050509b}.
\newblock {\em The Astrophysical Journal}, 630(2):L165--L168, aug 2005.

\bibitem{early_warning}
Kipp Cannon, Romain Cariou, et~al.
\newblock {TOWARD} {EARLY}-{WARNING} {DETECTION} {OF} {GRAVITATIONAL} {WAVES}
  {FROM} {COMPACT} {BINARY} {COALESCENCE}.
\newblock {\em The Astrophysical Journal}, 748(2):136, Mar 2012.

\bibitem{prompt_flash}
Ehud Nakar.
\newblock Short-hard gamma-ray bursts.
\newblock {\em Physics Reports}, 442(1):166 -- 236, 2007.
\newblock The Hans Bethe Centennial Volume 1906-2006.

\bibitem{NSBH_EMemission}
Sean~T. McWilliams and Janna Levin.
\newblock {ELECTROMAGNETIC EXTRACTION OF ENERGY FROM BLACK-HOLE–NEUTRON-STAR
  BINARIES}.
\newblock {\em The Astrophysical Journal}, 742(2):90, Nov 2011.

\bibitem{resonant_shattering}
David Tsang, Jocelyn~S. Read, Tanja Hinderer, Anthony~L. Piro, and Ruxandra
  Bondarescu.
\newblock {Resonant Shattering of Neutron Star Crusts}.
\newblock {\em Phys. Rev. Lett.}, 108:011102, Jan 2012.

\bibitem{magnetic_interaction_in_BNS}
Anthony~L. Piro.
\newblock {MAGNETIC} {INTERACTIONS} {IN} {COALESCING} {NEUTRON} {STAR}
  {BINARIES}.
\newblock {\em The Astrophysical Journal}, 755(1):80, jul 2012.

\bibitem{FRB_BH_battery}
Chiara M.~F. Mingarelli, Janna Levin, and T.~Joseph~W. Lazio.
\newblock {FAST} {RADIO} {BURSTS} {AND} {RADIO} {TRANSIENTS} {FROM} {BLACK}
  {HOLE} {BATTERIES}.
\newblock {\em The Astrophysical Journal}, 814(2):L20, nov 2015.

\bibitem{bayestar}
Leo~P. Singer and Larry~R. Price.
\newblock {Rapid Bayesian position reconstruction for gravitational-wave
  transients}.
\newblock {\em Phys. Rev. D}, 93:024013, Jan 2016.

\bibitem{bayestar3d}
Leo~P. Singer, Hsin-Yu Chen, et~al.
\newblock {GOING} {THE} {DISTANCE}: {MAPPING} {HOST} {GALAXIES} {OF} {LIGO}
  {AND} {VIRGO} {SOURCES} {IN} {THREE} {DIMENSIONS} {USING} {LOCAL}
  {COSMOGRAPHY} {AND} {TARGETED} {FOLLOW}-{UP}.
\newblock {\em The Astrophysical Journal}, 829(1):L15, sep 2016.

\bibitem{LALInference}
J.~Veitch, V.~Raymond, B.~Farr, W.~Farr, P.~Graff, S.~Vitale, B.~Aylott,
  K.~Blackburn, N.~Christensen, M.~Coughlin, et~al.
\newblock {Parameter estimation for compact binaries with ground-based
  gravitational-wave observations using the LALInference software library}.
\newblock {\em Physical Review D}, 91(4), Feb 2015.

\bibitem{url_sphradiometer}
Kipp Cannon and Takuya Tsutsui.
\newblock sphradiometer-0.3.0, October 2020.
\newblock 10.5281/zenodo.4276523
  (\url{https://doi.org/10.5281/zenodo.4276523}).

\bibitem{rapid_pseudo_bayestar}
Hsin-Yu Chen and Daniel~E. Holz.
\newblock {Facilitating Follow-up of LIGO–Virgo Events Using Rapid Sky
  Localization}.
\newblock {\em The Astrophysical Journal}, 840(2):88, May 2017.

\bibitem{KAGRA1}
Kentaro Somiya.
\newblock {Detector configuration of KAGRA–the Japanese cryogenic
  gravitational-wave detector}.
\newblock {\em Classical and Quantum Gravity}, 29(12):124007, Jun 2012.

\bibitem{KAGRA2}
Yoichi Aso, Yuta Michimura, Kentaro Somiya, Masaki Ando, Osamu Miyakawa,
  Takanori Sekiguchi, Daisuke Tatsumi, and Hiroaki Yamamoto.
\newblock {Interferometer design of the KAGRA gravitational wave detector}.
\newblock {\em Phys. Rev. D}, 88:043007, Aug 2013.

\bibitem{maggiore1}
M.~Maggiore.
\newblock {\em {Gravitational Waves: Volume 1: Theory and Experiments}}.
\newblock Gravitational Waves. OUP Oxford, 2008.

\bibitem{Sutton}
Patrick~J Sutton, Gareth Jones, Shourov Chatterji, Peter Kalmus, Isabel Leonor,
  Stephen Poprocki, Jameson Rollins, Antony Searle, Leo Stein, Massimo Tinto,
  and Michal Was.
\newblock {X-Pipeline: an analysis package for autonomous gravitational-wave
  burst searches}.
\newblock {\em New Journal of Physics}, 12(5):053034, may 2010.

\bibitem{Kipp}
Kipp~C. Cannon.
\newblock {Efficient algorithm for computing the time-resolved full-sky cross
  power in an interferometer with omnidirectional elements}.
\newblock {\em Phys. Rev. D}, 75:123003, Jun 2007.

\bibitem{healpy}
Andrea Zonca, Leo Singer, Daniel Lenz, Martin Reinecke, Cyrille Rosset, Eric
  Hivon, and Krzysztof Gorski.
\newblock {healpy: equal area pixelization and spherical harmonics transforms
  for data on the sphere in Python}.
\newblock {\em Journal of Open Source Software}, 4(35):1298, March 2019.

\bibitem{healpix}
K.~M. {G{\'o}rski}, E.~{Hivon}, A.~J. {Banday}, B.~D. {Wandelt}, F.~K.
  {Hansen}, M.~{Reinecke}, and M.~{Bartelmann}.
\newblock {HEALPix: A Framework for High-Resolution Discretization and Fast
  Analysis of Data Distributed on the Sphere}.
\newblock {\em \apj}, 622:759--771, April 2005.

\bibitem{gstlal1}
Cody Messick, Kent Blackburn, et~al.
\newblock {Analysis framework for the prompt discovery of compact binary
  mergers in gravitational-wave data}.
\newblock {\em Phys. Rev. D}, 95:042001, Feb 2017.

\bibitem{gstlal2}
Surabhi Sachdev, Sarah Caudill, et~al.
\newblock {The GstLAL Search Analysis Methods for Compact Binary Mergers in
  Advanced LIGO's Second and Advanced Virgo's First Observing Runs}, 2019.

\bibitem{url_ligo_skymap}
{ligo.skymap}.
\newblock \url{https://lscsoft.docs.ligo.org/ligo.skymap/index.html}.

\bibitem{CTA}
I~Bartos, T~Di~Girolamo, J~R Gair, M~Hendry, I~S Heng, T~B Humensky, S~Márka,
  Z~Márka, C~Messenger, R~Mukherjee, et~al.
\newblock {Strategies for the follow-up of gravitational wave transients with
  the Cherenkov Telescope Array}.
\newblock {\em Monthly Notices of the Royal Astronomical Society},
  477(1):639–647, Mar 2018.

\bibitem{open_data_appreciate}
The LIGO~Scientific Collaboration, the Virgo~Collaboration, et~al.
\newblock {Open data from the first and second observing runs of Advanced LIGO
  and Advanced Virgo}, 2019.

\bibitem{Jolien}
Jolien D. E. (Jolien Donald~Earl) Creighton and Warren~G. Anderson.
\newblock {\em Gravitational-wave physics and astronomy : an introduction to
  theory, experiment and data analysis}.
\newblock Wiley-VCH, 2011.

\end{thebibliography}

\end{document}